\documentclass[%
aps,
pra,
amsmath,amssymb,
reprint,%
superscriptaddress,
floatfix
]{revtex4-2}

\usepackage{dcolumn}%
\usepackage{bm}%
\usepackage[utf8]{inputenc}
\usepackage[T1]{fontenc}
\usepackage{mathptmx}
\usepackage{etoolbox}
\usepackage{physics}
\usepackage{graphicx}%
\graphicspath{./ }
\usepackage{amsmath}
\usepackage[dvipsnames]{xcolor}
\usepackage[colorlinks=true,
citecolor=blue,
linkcolor=magenta]{hyperref}

\begin{document}

\title{Multi-watt 1 GHz single-cycle frequency combs}
\author{Yanyan Zhang}
 \affiliation{Northwestern Polytechnical University, School of Artificial Intelligence, Optics and Electronics, Xi’an 710072, China}
 \affiliation{Research \& Development Institute of Northwestern Polytechnical University in Shenzhen, Shenzhen 518063, China}
 \author{Sida Xing}
 \email{xingsida@siom.ac.cn}
 \affiliation{Shanghai Institute of Optics and Fine Mechanics, Chinese Academy of Sciences, Shanghai 201800, China}

\date{\today}%

\begin{abstract}
Single-cycle optical pulses offer a strong carrier-envelope-offset (CEO) dependent electric field and the highest peak intensity for a given pulse energy. Absence of demonstrated GHz single-cycle lasers constrains exploration of single/sub-cycle dynamics at this repetition rate. By leveraging fiber soliton effects and suppressing higher-order dispersion, we achieve single-cycle pulse generation at a 1 GHz repetition rate in an all-fiber format. The laser produces 7.1 fs (1.1-cycle) pulses with 1.8 W average power, centered around 1970 nm. Temporal characterization shows 60\% of the pulse energy is concentrated in the pulse center, yielding a peak power of 110 kW. The seed laser demonstrates a 43 dB signal-to-noise ratio for the CEO frequency, facilitating comb stabilization and CEO control. Our model, which matches experimental observations, identifies conditions for achieving single-cycle duration and predicts scalability to a 2 GHz repetition rate. This work presents the first GHz single-cycle source. We envision these advances will drive studies in single/sub-cycle light-matter interaction, spectroscopy, microscopy, and CEO-sensitive nonlinear optics.

\end{abstract}
\maketitle
\section{Introduction}
Single-cycle optical pulses offer significant advancements in both field and intensity-related applications by enabling precise control of peak electric fields. This capability is crucial for sub-cycle manipulation of tunneling effects at the atomic scale \cite{Garg2020}, attosecond optical field sampling \cite{Ludwig2020, Bionta2021, Zimin2021}, creating attosecond electron bunches \cite{Kruger2011, Garg2016}, and probing valence electron motion \cite{Morimoto2021, Goulielmakis2010}. For peak intensity applications, their high peak power and broad in-phase spectrum are ideal for creating frequency combs via coherent supercontinuum generation \cite{Dudley2004} or intra-pulse difference frequency generation (IP-DFG) \cite{Kowligy2019, Steinleitner2022}. High repetition rate (> 100 MHz) frequency comb sources offer faster data acquisition and better absorption features per unit time, essential for studying rapidly varying processes like combustion dynamics and laser wavelength drifts \cite{Hoghooghi2023CompleteRO, Long2023, Giorgetta2010}. Ideally, a single-cycle pulse concentrates all energy in its center burst, enhancing power efficiency, reducing background noise from satellite pulses, and suppressing thermal effects on samples \cite{Garg2020}. Additionally, the aforementioned applications require nanojoule pulses with carrier-envelope-offset (CEO) control or stabilization, adding extra demands on the sources. The limited pulse energy of GHz repetition rate sources, combined with the need for single-cycle and high pulse quality, presents formidable challenges for construction, characterization, and operation.

Since the invention of the femtosecond laser in 1974 \cite{Shank1974}, various techniques have been developed to generate sub-two-cycle pulses at different pulse energies and repetition rates \cite{Yamane2003, Kobayashi2007, Couairon2005, Fuji2007, Krauss2010, Yoshioka2019, ChenBH2019, Hoghooghi2022, Steinleitner2022, Kowalczyk2023}. Ti:sapphire lasers \cite{Fuji2007, Nogueira2008}, optical parametric amplifiers/oscillators \cite{McCracken2015}, and Yb-doped fiber lasers \cite{Labaye2022} can directly emit nanojoule-level few-cycle pulses at GHz repetition rates, making them promising candidates for single-cycle frequency comb sources. However, achieving Watt-level average power for GHz sources necessitates amplification, and further spectrum broadening is required for single-cycle compression. While fibers facilitate extended interaction lengths for amplification and spectrum broadening, suppressing higher-order dispersion (HOD) remains a significant challenge. Pulse compression often involves complex dispersion management \cite{Lesko2021} or the synthesis of pulses at multiple wavelengths \cite{Krauss2010, Manzoni2015}. A fundamental solution involves minimizing the spatial accumulation of HOD, which is typically constrained by fiber amplifier efficiency. This challenge was addressed using heavily-doped Thulium-doped fiber amplifiers (TDFA), enabling 100 MHz single-cycle frequency combs \cite{Xing2021}. However, the feasibility of this approach at GHz repetition rates is uncertain. In addition, without an experimentally validated model, predicting the scalability of this method is challenging.

\begin{figure*}[!hpbt]
    \centering
    \includegraphics[width=.8\linewidth]{ 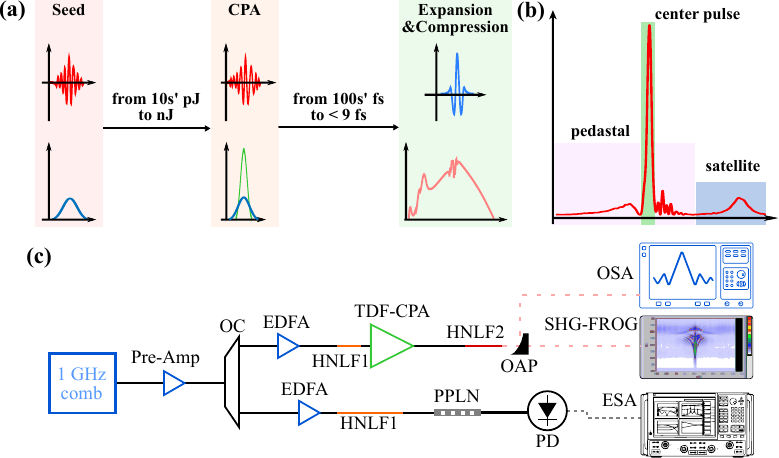}
	\caption{(a) The general principle behind this experiment with the major objectives in each stage. An 1970 nm, 80 fs Raman soliton from 1 GHz seed comb gets amplified from about 70 pJ (70 mW) to 2 nJ (2 W) in a chirped pulse amplifier. The targeted pulse duration is less than 9 fs after a dual-stage compressor. At this pulse duration, more than one octave spectrum appears at the same time frame. (b) Ideally, a compressed pulse should constrain all energy in pulse center. Due to imperfect dispersion compensation, the pulse energy is distributed among pulse center, pedestal and satellite pulses. The center pulse energy is vital for reaching single-optical-cycle domain. It also contributes to nearly all electric field/power intensity in single-cycle pulse applications. (c) The "2-branch" experimental implementation, where seed laser is equally split into two ports after pre-amplification. OC: optical coupler; EDFA: Erbium-doped fiber amplifier; HNLF: highly nonlinear fiber; TDF-CPA: Thulium-doped fiber chirped pulse amplifier; PPLN: periodically-poled lithium niobate; OSA: optical spectrum analyzer; SHG-FROG: second-harmonic-generation frequency-resolved-optical-gating; ESA: electrical spectrum analyzer; PD: photodiode; OAP: off-axis parabolic mirror. }
    \label{fig:principle}
\end{figure*}

To the best of our knowledge, we present the first demonstration of single-cycle pulse generation at a 1 GHz repetition rate. Our device features an all-fiber configuration with turnkey operation capability. A commercial 1 GHz frequency comb at 1560 nm is split into two ports. One port experiences soliton self-frequency shift to provide a chirp-free seed at 2 µm. After pre-chirping to normal dispersion, a Thulium-doped fiber amplifier (TDFA) delivers 2 W average power with over 40\% efficiency. The amplified pulse then passes through two stages of self-compression. The first stage compensates residual net-normal dispersion and boosts the peak power to approximately 20 kW. This pulse then enters a 7.7 cm highly-nonlinear fiber (HNLF), where it is further compressed. An all-reflective second-harmonic-generation frequency-resolved-optical-gating (SHG-FROG) measures the pulse duration to be about 7.1 fs (1.1 optical cycles) with nearly 60\% energy in the pulse center. After the HNLF, the output power of the single-cycle laser is 1.8 W, reduced due to splicing and facet losses. At 1.8 nJ pulse energy, the peak power is approximately 110 kW. The other port retrieves the carrier-envelope-offset (CEO) frequency, \( f_{ceo} \), with a 43 dB signal-to-noise ratio (SNR) at 300 kHz resolution bandwidth (RBW). This high-quality \( f_{ceo} \) is essential for comb stabilization and CEO control in future applications. Our numerical model aligns well with experimental data, including the single-cycle pulse spectrum, temporal distribution, and optimal HNLF length. Based on the model, a scaling law predicts the feasibility of doubling the repetition rate or average power while maintaining single-optical-cycle duration.

\section{Experimental setup}

Figure \ref{fig:principle}(a) shows the top-level principle of our experiment. A transform-limited soliton is generated by frequency shift a commercial 1 GHz frequency comb at 1560 nm. The Raman soliton self-frequency shift process moves the soliton to 1970 nm with about 70 pJ pulse energy. In a chirped pulse amplifier (CPA) based-on thulium-doped fibers (TDF), the seed is amplified by approximately 30 times, reaching 2 W average power (2 nJ pulse energy). The total length of the TDF is about 60 cm. The fiber stretcher provides strong normal dispersion to avoid nonlinear effects inside amplifier, whose amplified spontaneous emission can evolve in nonlinear process and degrades pulse coherence. At the end, self-compression provides spectrum broadening and pulse compression in the same fiber segment. We performs a dual-stage self-compression: the first PM1550 fiber stage features residual dispersion compensation and initial compression; the second stage utilize an HNLF to ensure the a soliton order close to 1 for single cycle pulse. 

\begin{figure*}[htbp]
       \includegraphics[width=.9\linewidth]{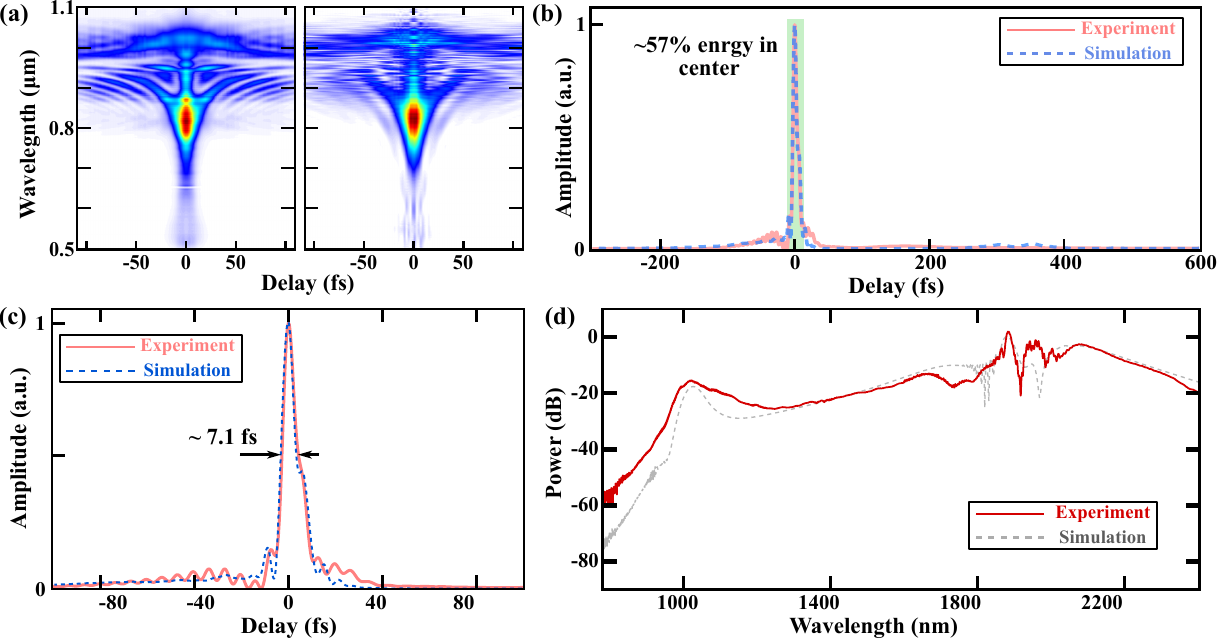}
        \centering
        \caption{ (a) Experiment (left) and reconstructed spectrogram over $\pm100$ fs displayed in linear scale. The FROG reconstruction error is 0.8\%. Despite the fast droping response of silicon detector array over 1000 nm, the spectrogram clearly indicate over an octave of SHG spectrum(500 nm to 1100 nm) is in-phase. (b) The retrieved pulse temporal distribution over the -300 fs to 600 fs delay range. A small pedestal with little satellite pulse is recorded. For better comparison, experimental (red) and simulated (blue) pulses are superimposed. The center pulse is highlighted in green and has 57\% of total pulse energy. (c) Retrieved (red) and simulated (blue) pulse temporal distribution over $\pm$100 fs, indicating 7.1 fs pulse duration. The simulation can provide a good matching of pulse including its pedestal; (d) Experimental (red) and simulated (grey) pulse spectrum covering 600 nm to beyond 2400 nm. Note that the peak around 1 µm is due to dispersive wave and is independent from the 2 µm pulse.}
        \label{fig:pulse_measurement}
\end{figure*}

When a pulse propagates inside optical fibers, the nonlinear chirp induced by self-phase modulation can only compensate group velocity dispersion (GVD, or 2nd dispersion) near the middle of the pulse. As a result, HODs cannot be removed by self-compression. Actually, most optical fibers has the same sign of third-order dispersion (TOD), increasing the total TOD even during self-compression. The leading and trailing edges of the compressed pulse create the pedestals; and higher order dispersion causes satellite pulses [Fig. \ref{fig:principle}(b)]. The pedestal and satellites "steal" energy from pulse center and lower the actual field strength/pulse intensity. In optical field-related applications such as sub-cycle electron dynamics \cite{Garg2020}, the single optical cycle electric field refers to the field near the center of pulses, when they are highly temporally confined. For optical intensity-related application such as IP-DFG \cite{Kowligy2019, Steinleitner2022}, the power in pulse center is normally referred as peak power and solely responsible for the IP-DFG process. Less split of energy from center also helps single cycle pulse formation - this process tends to prevent formation of single-optical-cycle pulses. Therefore, high quality pulse is vital for both single-cycle sources and their associated applications. 

Figure \ref{fig:principle}(c) sketches the experimental implementation. To avoid polarization effects on frequency comb stabilization and improve the efficiency of self-compression, our setup features an all-polarization-maintaining fiber configuration. The seed laser splits to two branches for single-cycle frequency frequency comb and f-2f interferometry, respectively. In the f-2f branch, the octave-spanning spectrum from HNLF1 is frequency doubled on a periodically-poled lithium niobate crystal and generate the $f_{ceo}$ node after a photodetector. The residual light can provide the beat node with a reference laser. The same process generates the 2 µm seed soliton in the other branch. A core-pumped double-cladding TDFA amplifies the pulse to 2 W average power (2 nJ) within 0.75 m of total fiber length. The dual-stage compressor consists of 85 cm PM1550 and 7.7 cm of HNLF2. The HNLF2 has a GVD of -13 ps\textsuperscript{2}/km and nonlinear parameter, $\gamma$ = 3.1 (W-km)\textsuperscript{-1}. After compression, an off-axis-parabolic mirror collimator send the light for spectral and temporal characterization. 

\section{Results and discussions}

\begin{figure}[!ht]
    \centering
    \includegraphics[width=.85\linewidth]{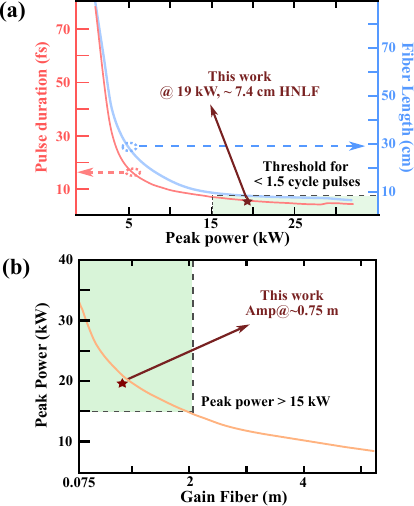}
    \caption{(a) The minimum pulse duration after self-compression (Left-axis) and the optimal fiber length (Right-axis) are plotted as a function of pulse energy of amplified 80 fs pulse. For better interpretation, the pulse energy is transferred to peak power. (b) Achievable peak power as a function of gain fiber length. The maximum gain fiber length is about 2 m, showing the scalability to double the pulse energy or repetition rate. }
    \label{fig:scalability}
\end{figure}

With the all-reflective SHG-FROG, we record the spectrogram of output pulse in Fig. \ref{fig:pulse_measurement}(a) zoomed in to $\pm$100 fs. Despite the drop of silicon detector array response over 1100 nm, the spectrogram shows that more then an octave spectrum (below 500 nm to 1100 nm) is in phase. The FROG reconstruction error is 0.8\% based on equation (A.2) in \cite{Rhodes2013}. The FROG scan indicates near-zero power exists beyond -300 fs and +600 fs. Thus, we only plot FROG retrieved pulse from -300 fs to +600 fs in Fig. \ref{fig:pulse_measurement}(b). Integrating over the time delay, the center pulse persists over 57\% of total pulse energy. As shown in Fig. \ref{fig:pulse_measurement}(c), the reconstructed pulse duration is 7.1 fs (1.1 optical-cycle at 1970 nm) and exhibit very low pedestal, leading to a peak power of about 110 kW. We use two optical spectrum analyzers to record the output spectrum over 600 nm to 2400 nm. Displayed in Fig. \ref{fig:pulse_measurement}(d), the output pulse spans over two octaves. The peak at around 1000 nm results from dispersive-wave generation and is an independent pulse experiencing normal dispersion. 

We generate fundamental order soliton through self-frequency shift process, which can be described with a chirp-free sech\textsuperscript{2} function. Using Raman soliton as initial conditions allows numerical solver predicts experimental results with good precision \cite{xing2018linearly}. Experimentally, the soliton duration is about 80 fs and energy is 70 pJ. After HNLF1 [\ref{fig:pulse_measurement}(c)], the soliton numerically propagates through the entire laser using generalized nonlinear Schrodinger equation (GNLSE) \cite{Dudley2006}. We superimpose the simulated pulse temporal profile and spectrum with experimental data as dashed curve in Fig. \ref{fig:pulse_measurement}(b)-(d). The good agreement enables scaling of single-cycle source conditions with high confidence. 

We relax the definition of "single cycle pulses" as pulses with duration less than 9.8 fs (1.5 optical cycles at 2 µm) in this section. A backward numerical calculation from HNLF2 is performed to evaluate the pump condition to reach single-optical-cycle. Before entering the HNLF2, we assume a perfect pulse without satellites or pedestal. Experimentally, this ideal pulse is equivalent to the pulse center as indicated in \ref{fig:principle}(b). Assuming the pulse duration is the same as seed pulse (80 fs), the achievable pulse duration after HNLF2 decreases at higher pulse energy (i.e. higher peak power) µm, as the red curve in \ref{fig:scalability}(a). The pulse energy is converted to pulse peak power to ease later calculations. The threshold for single-cycle pulse generation is 1.3 nJ energy with maximum 8 cm HNLF2. Below the threshold energy, pulses experience excessive degradation from HODs, excluding it from reaching single-optical-cycle. The corresponding length when HNLF2 compresses the pulse to minimum duration is plotted in the blue curve of Fig. \ref{fig:scalability}(a). It is the combined effort of TDFA and first stage of compression that provides the necessary peak power to drive self-compression inside HNLF2. With this condition, we numerically trace back to the fist-stage of self-compression and the amplifier. After amplifying to 2 nJ , we estimate the impact of amplifier length to the achievable peak power [Fig. \ref{fig:scalability}(b)]. Notably, the results indicate feasibility to double the TDFA length. With doubled TDFA length, 2 GHz single cycle sources are plausible with same gain fiber. Longer TDFA could double the pulse energy of 1 GHz single cycle pulses. Since self-compression is a peak-intensity-related process, this seemingly rough estimation provides good matching with our later experiment [Fig. \ref{fig:scalability}(a) and (b)]. 

\begin{figure}[!ht]
    \centering
    \includegraphics[width=.7\linewidth]{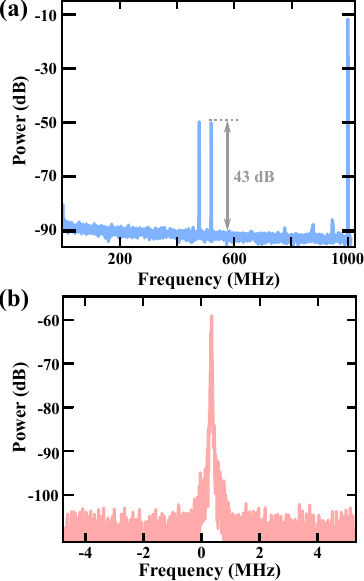}
    \caption{(a). The $f_{ceo}$ at 300 kHz resolution bandwidth, showing 43 dB signal-to-noise ratio. (b) The $f_{ceo}$ (offset to around 0 MHz) beat node at 10 kHz resolution bandwidth with about 40 kHz free-running linewidth. }
    \label{fig:fceo}
\end{figure}

The f-2f branch utilizes the same HNLF1 for octave-spanning frequency generation [\ref{fig:principle}(c)]. The f-2f interferometry also gives a chance to check the coherence of the 2 µm seed soliton. Figure \ref{fig:fceo}(a) and (b) shows the recorded f\textsubscript{ceo} nodes at at 300 kHz RBW and 10 KHz RBW, respectively. Its SNR is about 42 dB at 300 kHz RBW while the free-running linewidth is about 40 kHz. Previous frequency comb stabilization report on the same seed laser suggests that our current f\textsubscript{ceo} can support CEO stabilization with low phase noise\cite{Lesko2020_1GHz}. This single-cycle laser can also be carrier-envelope-phase-stabilized (CEP-stabilized) following the approach in \cite{Okubo2018}. 

\section{Conclusion and future work}
To conclude, we presented a single-cycle laser centered at 1970 nm, operating at a 1 GHz repetition rate with 2 W average power. Characterized using an all-reflective FROG, the output pulse duration is 7.1 fs, equivalent to 1.1 optical cycles. With approximately 60\% of the energy concentrated in the pulse center, the peak power is calculated to be 110 kW. To the best of our knowledge, this is the first demonstration of single-cycle pulse generation at a 1 GHz repetition rate. The $f_{ceo}$ node, with a 43 dB SNR, allows for the stabilization of each comb line to RF standards and CEO phase control. This single demonstration achieves single-cycle generation, high pulse quality, 2 W average power, over 40\% pump efficiency, CEP control capability, an all-fiber configuration, and turnkey operation. Our validated numerical model outlines the scaling rules of our device and predicts the feasibility of reaching a 2 GHz repetition rate. Further rigorous studies on the pulse dynamics, especially sub-cycle dynamics, are underway. Additionally, more general design rules to fully exploit the capabilities of self-compression in single-cycle sources are being developed.
\\
\section*{Supplementary information}
See the supplementary material for supporting content.
\\
\section*{Declaration}
The mention of specific companies, products, or trade names does not constitute an endorsement by SIOM. The authors declare no conflict of interest.
\\
\section*{Data Availability}
All data needed to evaluate the conclusions in the paper are present in the paper and/or the Supplementary Materials. Additional data and codes are available from the corresponding authors upon reasonable request.
\\
\bibliography{2octave_MIR}

\end{document}